\documentclass[aps,showpacs]{revtex4}
\usepackage{graphicx}         

\begin{document}


\newlength{\defbaselineskip}
\setlength{\defbaselineskip}{\baselineskip}
\newcommand{\setlinespacing}[1]
           {\setlength{\baselineskip}{#1 \defbaselineskip}}
\newcommand{\doublespacing}{\setlength{\baselineskip}
                           {2.0 \defbaselineskip}}
\newcommand{\singlespacing}{\setlength{\baselineskip}{\defbaselineskip}}
\newcommand{\1}{{\mbox{\hspace{1pt}}}}
\newcommand{\tab}{\hbox{\hspace{3ex}}}

\newcommand{\eqn}[1]{Eq.~(\ref{#1})}
\newcommand{\Eqn}[1]{Equation~(\ref{#1})}
\newcommand{\fig}[1]{Fig.~\ref{#1}}
\newcommand{\Fig}[1]{Figure~\ref{#1}}
\newcommand{\secn}[1]{Section~\ref{#1}}
\newcommand{\Secn}[1]{Section~\ref{#1}}
\newcommand{\defn}[1]{definition~\ref{#1}}
\newcommand{\Defn}[1]{Definition~\ref{#1}}

\newcommand{\abs}[1]{|#1|}
\newcommand{\realpart}[1]{\,\mathrm{Re}\!\left\{#1\right\}}
\newcommand{\equivalent}{\Leftrightarrow}
\newcommand{\hence}{\Rightarrow}
\newcommand{\w}{\omega}
\newcommand{\scr}[1]{{\scriptscriptstyle{#1}}}
\newcommand{\sub}[1]{_{\mathrm{#1}}}
\newcommand{\supscr}[1]{^{\mathrm{#1}}}
\newcommand{\e}[1]{\hat{e}_{#1}}
\newcommand{\E}[1]{\times 10^{#1}}
\newcommand{\lsim}{\;$\raisebox{-0.5ex}{$\stackrel{<}{\scriptstyle{\sim}}$}$\;}
\newcommand{\gsim}{\;$\raisebox{-0.5ex}{$\stackrel{>}{\scriptstyle{\sim}}$}$\;}

\newcommand{\unitmatrix}{{\mathbf 1}}

\newcommand{\dnd}[3]{\frac{\partial^{#1} #2}{\partial #3^{#1}}}
\newcommand{\ddd}[2]{\dnd{2}{#1}{#2}}
\newcommand{\dd}[2]{\frac{\partial #1}{\partial #2}}
\newcommand{\ddt}[1]{\dd{#1}{t}}
\newcommand{\ddx}[1]{\dd{#1}{x}}
\newcommand{\ddy}[1]{\dd{#1}{y}}
\newcommand{\ddz}[1]{\dd{#1}{z}}

\newcommand{\V}[1]{\underline{#1}}
\newcommand{\grad}{\bigtriangledown}

\newcommand{\set}[1]{\{#1\}}
\newcommand{\R}{{\mathbb R}}
\newcommand{\C}{{\mathbb C}}

\newcommand{\units}[1]{\mathrm{\scriptstyle{#1}}}
\newcommand{\degrees}{^\circ}
\newcommand{\micron}{\mu\mathrm{m}}
\newcommand{\microns}{\mu\mathrm{m}}
\newcommand{\chem}[1]{$\mathrm{#1}$}

\newcommand{\state}[1]{|{\mathbf{#1}}\!\!>}
\newcommand{\dual}[1]{<\!\!{{\mathbf{#1}}}|}
\newcommand{\op}[1]{\mathbf{#1}}
\newcommand{\opdensity}[1]{{\mathbf{#1}}}
\newcommand{\ip}[2]{<\!\!{#1}|{#2}\!\!>}
\newcommand{\opip}[2]{\ip{\mathbf{#1}}{\mathbf{#2}}}
\newcommand{\T}{{}^{\dagger}}
\newcommand{\unitvec}[1]{\V{{\hat{#1}}}}

\newtheorem{example}{Example}

\title{Effects of electrical charging on the mechanical
Q of a fused silica disk}

\author{Michael~J.~Mortonson, Christophoros~C.~Vassiliou, 
David~J.~Ottaway, David~H.~Shoemaker, 
Gregory~M.~Harry\footnote{gharry@ligo.mit.edu}}
\affiliation{LIGO Laboratory, Massachusetts Institute of
Technology, Room~NW17-161, 175~Albany
Street, Cambridge, Massachusetts 02139, USA.
\vspace{2ex}LIGO-P020014-00-D}

\date{\today}

\begin{abstract}
We report on the effects of an electrical charge on mechanical loss of a fused silica disk.
A degradation of Q was seen that correlated with charge on the surface of the sample.  We
examine a number of models for charge damping, including eddy current damping and loss due to
polarization.  We conclude that rubbing friction between the sample and a piece
of dust attracted by the charged sample is the most likely explanation for the observed loss.
\end{abstract}


\maketitle
\section{Introduction}\label{Introduction}
Experimental efforts to measure gravitational waves have been ongoing
for over 40~years~\cite{Weber} and recently several interferometers~\cite{LIGO,GEO,TAMA}
have begun taking science data; no gravitational wave signals have been
observed to date.  Estimates of the strength
and rate of gravitational wave events are such that improvements in the sensitivity
of detectors would be richly rewarded in terms of event rates and information
extracted from the signals~\cite{300yrs}.  The sensitivity of these interferometers
will be limited by fundamental noise sources, with thermal noise from
the internal degrees of freedom in the mirrors setting the limit at the frequency
of highest sensitivity.
Any increase beyond what is currently anticipated for thermal noise, or 
additional noise sources beyond what is expected~\cite{whitepaper}, will reduce 
the sensitivity of advanced interferometers.

Thermal noise is generally studied in the laboratory indirectly, through
application of the fluctuation-dissipation theorem~\cite{callen}, which
says that thermal noise can be predicted from the loss in the system.  
Studying the mechanical loss of optics is much simpler in the
laboratory than direct measurement of thermal noise.  Loss can be characterized
by the quality factor, $Q$, of a resonant mode, with higher Q's resulting in 
lower thermal noise.  A variety of mechanisms are thought to play a role in
introducing mechanical loss in a suspended gravitational-wave detector optic:
intrinsic losses in the mirror material, in the attachments to the 
suspension, in the dielectric optical coating~\cite{coating1,coating2}, 
and through interactions with the environment through electromagnetic couplings.

The buildup of electric charge on interferometer optics has been observed
in LIGO.  Fused silica optics are known to become
charged and to increase their charge over many months~\cite{Mitrofanov2002}.  
Degradation in the Q of a fused silica suspension due 
to charging has also been observed in a pendulum~\cite{GlasgowQ} as well
as a torsional~\cite{Mitrofanov2000} mode.  The effect of charging on internal 
mode thermal noise has not been well studied.  

We have investigated the effect of charging on the mechanical Q of a normal
mode of a fused silica disk.  We observed noticable change in the ringdown
of the mode when the optic is charged in particular circumstances.  We present 
possible explanations for this change and discuss its relevance for interferometric 
gravitational wave detectors.

\section{Theory}
\label{section:theory}

There are a number of mechanisms by which charge on an optic could cause excess
loss and thereby higher thermal noise.  We examine some of these mechanisms
to see which could cause excess mechanical loss in a laboratory setting.  This
list is not meant to be exhaustive.

\subsection{Eddy-current damping}
\label{section:eddycurrent}

One possible source of excess loss from charging is eddy-current damping between 
the charged sample and a nearby ground plane.  This is the mechanism which was 
suggested as an  explanation for the excess loss seen in the charged 
pendulum~\cite{GlasgowQ}.  We 
modeled the charged sample as a point mass placed near a ground plane infinite in extent.  
This ground plane could represent the metallic walls of a vacuum chamber, or a metal capacitor 
plate placed 
nearby the sample for exciting the normal modes~\cite{highQ}. The charge
was assumed to oscillate back and forth relative to the ground plane at 3~kHz, which 
is a typical normal mode frequency for laboratory experiments on small silica
samples.  This oscillating charge creates an oscillating field which 
induces currents in the ground plane.  These currents will suffer Ohmic losses in
the metallic plane.  This 
energy loss can be characterized by a limiting Q, 
\begin{equation}
\label{eq:Qlimit}
Q_{\mathrm{limit}} = 2 \pi E/ \Delta E_{\mathrm{cycle}},
\end{equation}
where $\Delta E_{\mathrm{cycle}}$ is the energy loss per cycle of oscillation,
and $E$ is the total elastic energy stored in the oscillation of the sample.  The energy loss 
$\Delta E_{\mathrm{cycle}}$ can be calculated using the equations of electrodynamics.

The surface charge on the ground plane is found to be 
\begin{equation}
\label{eq:surfacecharge}
\sigma(r,t) = -\frac{q (d + A \cos(2 \pi f t))}{2 \pi (r^2+d^2)^{3/2}},
\end{equation}  
where $q$ is the charge of the oscillating point mass, $d$ is the average distance
between the point mass and the ground plane, $A$ is the amplitude of the oscillation, 
and $f$ is the oscillation frequency. 
Since $\partial \sigma / \partial t = -\nabla \cdot J$, the surface 
current can be written
\begin{eqnarray}
J(r,t) = -\frac{1}{r} \int_0^r \frac{\partial}{\partial t} (r' \sigma(r',t)) dr'
\end{eqnarray}
Taking the power loss per area to be 
$\partial P / \partial S = \rho / 2\delta |{J(r,t)^2}|$, we get
\begin{eqnarray}
\Delta E = \frac{\pi \rho}{\delta} \int_0^{1/f} \int_0^r r' 
[J(r',t)]^2 dr dt
\end{eqnarray}
where $\rho$ is the resistivity and $\delta$ is the skin depth of the ground 
plane.  

For laboratory experiments, we used the following appropriate parameters;
an aluminum ground plane with $\rho = 2.7 \times 10^{-8}$~$\Omega\cdot$m and $\delta = 1.5 \times
10^{-3}$~m, a charge $q = 1.6 \times 10^{-9}$~C,
distance $d = 2 \times 10^{-3}$~m, $f = 3000$~Hz, and sample radius $r=3.8 \times 10^{-2}$~m. 
With these numbers, Eq.~(\ref{eq:Qlimit}) gives a limit to the Q of 
\begin{equation}
Q_{\mathrm{limit}} \sim 10^{20}.
\end{equation}
This Q is much higher than any modal Q seen in a material 
sample~\cite{sapphireQ,Glasgowsapphire,highQ,Numata}.
For normal mode ringdowns and interferometer thermal noise this mechanism was ruled out
as a relevant loss mechanism.

A second model is similar to the first but with a resistive wire between the 
plate and ground.  This could occur when the wires of the exciter are disconnected from the
high voltage used during excitation and grounded with a grounding cap.  The formula for 
surface charge on the on the plate is the same as 
Eq.~(\ref{eq:surfacecharge}).  By integrating this surface charge
over the area of the plate to get a total time-dependent charge, then 
differentiating with respect to time, a current that would flow 
through the wire, $I$, is obtained.  The value $\Delta E_{cycle}$ is found by using 
$P = I^2 R$  for power loss and integrating $P$ over one cycle.  For similar laboratory
values as above, with a wire resistance $R=1$~$\Omega$, a $Q_{\mathrm{limit}}$ of 
$\sim 10^{18}$ is obtained.  This is similarly higher than any reasonable material
Q, and thus we can conclude that loss due to induced currents flowing in a wire
between a plate and ground 
also has a  negligible effect on thermal noise.

\subsection{Polarization losses}
\label{section:polarization}

Another possible source of damping associated with surface charge is polarization 
loss.  As the charge on a sample moves relative to a ground plane,
the electric field inside the sample will change.  This changing electric field
inside the dielectric sample causes a changing polarization.  This changing 
polarization can be thought of as a current of bound charges which undergoes
loss as it flows.  The time-dependent electric field from a point charge of the 
models in Section~\ref{section:eddycurrent} was examined for its effect 
on a fused silica sample.  Since the sample is a dielectric medium, the changing
electric field causes the polarization of the sample to oscillate, creating 
``bound currents'',
\begin{equation}
J = \epsilon_0 \chi_e \frac{\partial E}{\partial t},
\end{equation}
where $E$ is the electric field, $\epsilon_0$ is the premittivity of free space,
and $\chi_e$ is the electric susceptibility of the sample material. 
The energy loss as a function of time is found by integrating $J^2$ over the volume of the 
sample and multiplying by the resistivity $\rho$.

The electric field used was the same as in Section~\ref{section:eddycurrent}, which
assumed a point charge next to an infinite ground plane.  Using those parameters for
the point charge, a frequency of 3~kHz, a silica sample $3.8 \times 10^{-2}$~m in radius, 
$2.5 \times 10^{-3}$~m thick,
and $2.6 \times 10^{-2}$~kg in mass, a silica resistivity of $2 \times 10^{12}~\Omega\cdot$m
and electric susceptibility of 2.8, and a distance between
the sample and the ground plane of $5 \times 10^{-3}$~m, a $Q$ of $10^{4}$ is obtained.
This $Q$ is a strong function of the distance between the charge and the ground plane, becoming
$2 \times 10^{7}$ at 3~cm.  With such a strong dependence, the approximations of the charge
as a point and the ground plane as infinite break down for a realistic laboratory setting.  The
resistivity of bulk silica is also not that well characterized.
This result does indicate that polarization loss in the body of a silica sample could be
an important loss mechanism when a ground plane is close to a charged optic.

A closely related loss mechanism is the polarization loss in a surface layer of a silica
sample.  This is the mechanism that was suggested for the result in~\cite{Mitrofanov2000}.
Using the same model as above, but only integrating over a surface region with a 
resistivity of $2 \times 10^{10}~\Omega\cdot$m,
the same susceptibility of 2.8, and a surface layers thickness of $10^{-5}$~m~\cite{amaldi},
results in a $Q$ of $ 2 \times 10^{10}$.  This mechanism shares the strong dependence on distance 
with the volume polarization loss, but also suffers from uncertainties in the properties of the 
surface layer.

\subsection{Electrostatic coupling to a lossy mechanical system}

There could be an electrostatic coupling between surface charge on a sample and a nearby
charged insulator.  This could occur if the insulation on wires in the exciter becomes charged
as well as the silica sample.  If motion of the surface charges can cause motion in the 
insulator, loss can occur either from rubbing friction between parts of a mechanical structure
or simply internal friction of the insulator material.  

This can be modeled as a coupled oscillator in 
which the oscillation of the sample induces vibration of the insulator.  The 
equations of motion can be solved to determine the energy lost to the insulator.  
Using a charge on both the sample and the insulator of $1.6 \times 10^{-9}$,
a sample mass of $0.26$~kg, an insulator mass of $2.6 \times 10^{-3}$~kg, a separation
of $2 \times 10^{-3}$~m, a frequency of $3000$~Hz, a spring constant between the 
insulator and mechanical ground of $3.4 \times 10^{7}$~N/m, and a loss angle for the insulator 
of $1 \times 10^{-3}$ results in a $Q$ of $2 \times 10^{9}$.  Certain parameters for this model,
notably the stiffness and loss angle of the insulator, are not well known for a laboratory 
setting.  Despite this, the high predicted Q compared to sample internal friction suggest
this should not be an important loss mechanism.

\subsection{Rubbing with dust particles}
\label{rubbing}

Another possible source of loss is rubbing between a dust particle and the sample.
This can be correlated with surface charge because a charged sample could attract
a charged dust particle, making the chance contact between dust and sample far more
likely when they are charged.  The friction force between the sample and a material
in contact with it can be written
\begin{equation}
F_{f} = - \mu N \mathbf{v}/|\mathbf{v}|,
\end{equation}
where $F_{f}$ is the frictional force, $\mu$ is the coefficient of friction, $N$ is 
the normal force between the two materials, and $\mathbf{v}$ is the velocity.  This velocity
dependence results in a force of constant magnitude in the opposite direction of the
relative velocity.  Solving the equation of motion for a system with this force 
results in a sinusoidal oscillation with a linearly decaying amplitude~\cite{coulomb};
\begin{equation}
x\left(t\right) = \left(A_{0} - \mu N t/\left( \pi^{2} f_0 m \right)\right) \sin 
\left(2 \pi f_0 t + \theta\right),
\label{eqn:linear}
\end{equation}
where $f_0$ is the frequency of the oscillation, $A_0$ is the initial amplitude of oscillation,
$m$ is the sample mass, and $\theta$ is an arbitrary phase.  Linear amplitude decay of this type
is known 
as Coulomb damping~\endnote{Coulomb damping is named for Charles Augustin de Coulomb who 
first  described it mathematically, although it is initially discussed in the literature by 
Aristotle.  Coulomb, of course, also did work with electrostatics where his 
name is frequently seen as well.  It is, however, a coincidence that the form of damping
described here is both called Coulomb damping and is associated with electric charge.}.
This distinctive form of decay allows
Coulomb damping to be distinguished easily from other sources of loss.
According to Eq.~(\ref{eqn:linear}), the amplitude of vibrations in a Coulomb damped 
system will decrease by $\mu N / \pi^2 f_0^2 m$ per cycle.  Thus the rate of decay should decrease
linearly as the normal force is reduced.

\section{Experiments}
\subsection{Method}
\label{section:method}

To test these sources of loss, we measured the mechanical quality factor, $Q$, 
of a charged fused silica disk. The disk was 76.2~mm in diameter by 2.5~mm thick, made from 
Corning 7980, Grade 0-A silica. We found the frequency of a normal mode of vibration,
excited this mode, and measured the ringdown.  From the decay time of this 
ringdown we were able to determine the effect of the charge on loss.  The normal mode measured
was the $n=1, \ell = 0$ mode, with a frequency of 4100~Hz.

The mode was excited using a comb capacitor~\cite{highQ,Abramovici} exciter.  
This consists of two wires wrapped side by side around an aluminum ground 
plane.  This exciter was placed close to the sample, typically about 1~cm, but
we were able to change this distance using a picomotor.  A DC voltage of 500~V
was placed on one of the two wires while the other was held at ground.  This
creates a diverging electric field near the exciter and inside the glass
dielectric.  An AC field with peak amplitude 500~V at the normal mode frequency
is then applied to the high voltage wire.  This AC field couples to the 
polarization in the glass to give a force on the sample at the normal mode
frequency.  The electric field and the exciter can interact with any charge
on the sample.

\begin{figure}
\includegraphics[width=12cm]{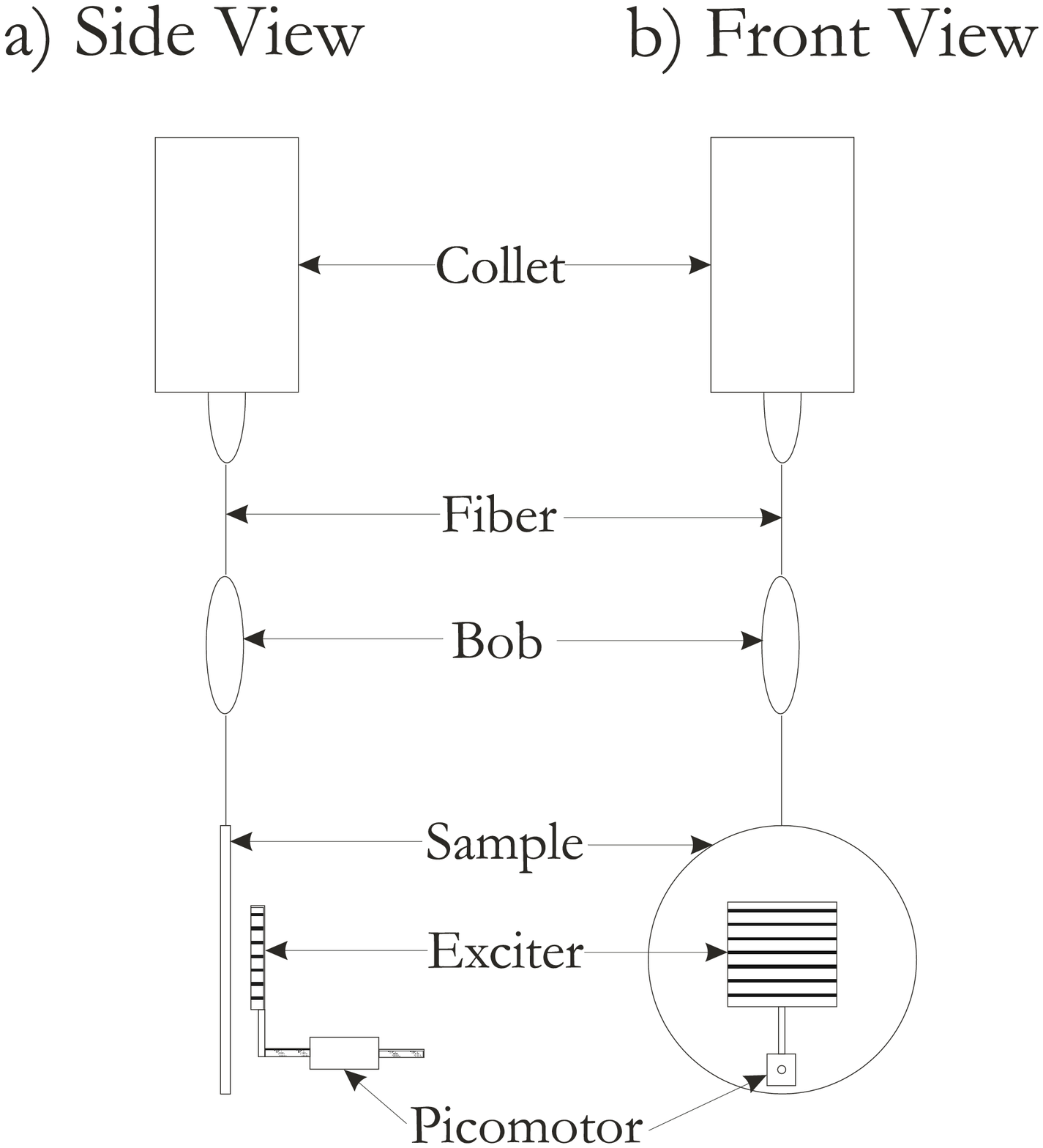}\vspace{2ex}
\caption{The setup used for these experiments.  The sample is suspended
below a thin fiber of silica welded to a massive bob which itself is supended
below a silica fiber welded to a silica mass held in a collet.  Next to the
sample is an exciter made from two wires wrapped around a ground plane.  The
exciter can be moved relative to the sample by the picomotor.  The oscillations
in the sample are read out using stress polarimetry with a polarized HeNe laser
as the probe.
}\label{fig:setup}
\end{figure}

The test sample was suspended by a monolithic, fused silica suspension of thin
fibers and a single isolation bob~\cite{highQ}.  The suspension is held on top by a collet
attached to an aluminum stand.  The monolithic fiber-bob suspension keeps excess
loss from recoil damping or rubbing at interfaces from affecting the Q measurement.
The entire setup is contained within a vacuum bell jar which is pumped down to at
least $10^{-3}$~Pa and typically about $3 \times 10^{-4}$~Pa to avoid loss
from gas damping.  The experimental setup is shown in Figure~\ref{fig:setup}.
This experimental setup is similar to ones used in previous experiments and is more 
fully described there~\cite{highQ,Gretarsson,SUcoating}.

The sample's normal mode amplitude is read out versus time using a stress polarimeter.
A polarized HeNe laser is passed through the sample where stress induced 
birefringence changes the laser's polarization.  After passing 
through a $\lambda$/4 plate, the beam's polarization oscillates at the mode
frequency with an amplitude proportional to the mode amplitude.  This signal
is read out using a polarizing beamsplitter and two photodiodes.  The signal
is passed through a current-to-voltage amplifier and then heterodyned to about
0.3~Hz by a lock-in amplifier.  Finally, the data is passed to an analog-to-digital
converter and recorded on a PC.  

The data stored on computer are typically of the form of a damped sinusoid.  Most
sources of loss cause an exponential decay to occur in the mode amplitude, so the
data can be fit to
\begin{equation}
x\left(t\right) = e^{-t/\tau} \sin\left(2 \pi f_{\mathrm{demod}} t + \theta\right),
\end{equation}
where $\tau$ is the decay time, $f_{\mathrm{demod}}$ is the frequency after 
demodulation, and $\theta$ is an unimportant phase.  From this fit, the decay time
$\tau$ can be determined.  This characterizes the loss, reported as the dimensionless
value, 
\begin{equation}
Q = \pi f_{0} \tau,
\end{equation}
where $f_{0}$ is the normal mode frequency.
We also observed linear decay in the mode amplitude, where the data can be fit to
\begin{equation}
x\left(t\right) = \left(m t + A_{0}\right) \sin \left(2 \pi f_{\mathrm{demod}} t + 
\theta \right).
\end{equation}
This behaviour is characteristic of Coulomb damping~\cite{coulomb}.

To control the charge on the surface of the sample we used two techniques.  First,
an ionized-nitrogen spraygun, which could be used on the sample to either increase
or decrease the charge.  It is difficult to control exactly where on the sample's
surface the charge ends up, so for studies involving charge distribution we used a
small piece of silk cloth.  Gently rubbing the sample with the silk allows a charge to be
built up in the rubbed area.  This allowed us to create a high charge density region
in the center, for example, while leaving the edges of the sample with low charge.  
We also used the silk cloth to get greater charge densities across the
entire face of the sample.

The distance between the exciter and the sample could be controlled very precisely
using a picomotor.  The picomotor was measured to move at $1.7 \times 10^{-5}$~m/s, about
a millimeter per minute.  Thus, by simply measuring the time the picomotor was engaged, 
we could tell how far 
the exciter had moved to within a few tens of microns.  The zero position in distance
was defined by when the exciter touched the sample. This could be determined by 
observing when the low frequency pendulum mode oscillations of the sample stopped
due to contact with the exciter.

\subsection{Trials}
\label{section:experiments}

To test the effect of charge on thermal noise, we performed ringdown experiments.  
We changed various parameters, including surface charge density,
surface charge distribution, and distance between the exciter and the sample, to
determine which, if any, of the phenomenon described in Section~\ref{section:theory}
was affecting the Q.  

\begin{figure}
\includegraphics[width=12cm]{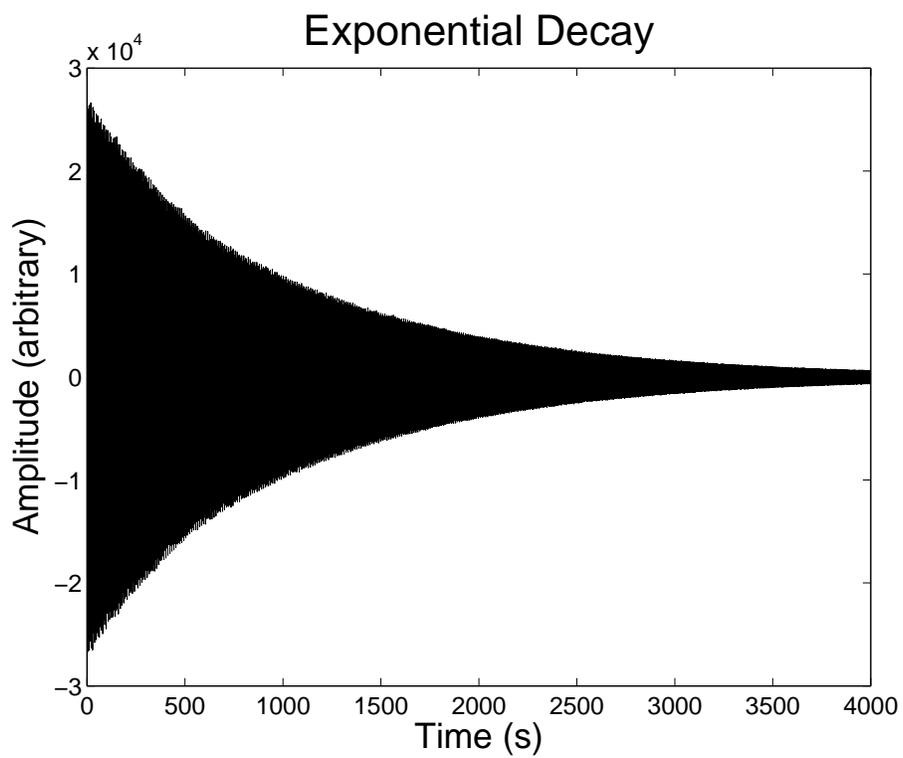}\vspace{2ex}
\caption{A typical modal ringdown when the sample is limited by 
the internal friction of the silica.  The envelope is exponential in 
shape.
}\label{fig:expringdown}
\end{figure}

To start, we collected Q data without any charge on the sample.  A typical ringdown
from these measurments is shown in Figure~\ref{fig:expringdown}.  The 
result is an
exponential ringdown with a quality factor of
\begin{equation}
Q = 13.3 \times 10^{6}.
\end{equation}
This measured Q showed no dependence on the distance between the exciter and the sample,
including down to separations below 300~$\mu$m (see below).

We measured the Q under a variety of charging conditions.  The sample was charged using the ionized nitrogen spraygun, to give a uniform charge 
on both sides of the sample.  To obtain an approximate value for the surface charge 
density we used an Ion Systems Model 775 Periodic Verification System electrostatic 
fieldmeter.  This gave an indication of the sign of the charge, as well as providing a 
method of verifying whether the charge density before and after the experiment remained 
the same.  The charge density attainable was very low and thus, we rubbed the surface of 
the sample using a piece of silk cloth.  The electric field was measured to be 
$1.7 \times 10^{5}$~V/m at a distance of approximately 2.5~cm from the sample surface, 
approximately 5 times higher than attainable with the ion spray.  Assuming an infinite 
charge plane this converts to a minimum charge density of $1.5 \times 10^{-5}$~C/m$^{2}$.  
The ion spray was subsequently used to discharge the surface.  
Next we varied the distribution of the charge on the surface.  The center of the sample 
was charged with the silk cloth, to a minimum value of $9 \times 10^{-6}$~C/m$^{2}$, 
again using the infinite plane approximation.  
We then gave the sample a uniform charge and measured the Q while the exciter was left
with a DC voltage.  This was to test the effect of a static force acting between the sample
and the exciter structure.  

\begin{figure}
\includegraphics[width=12cm]{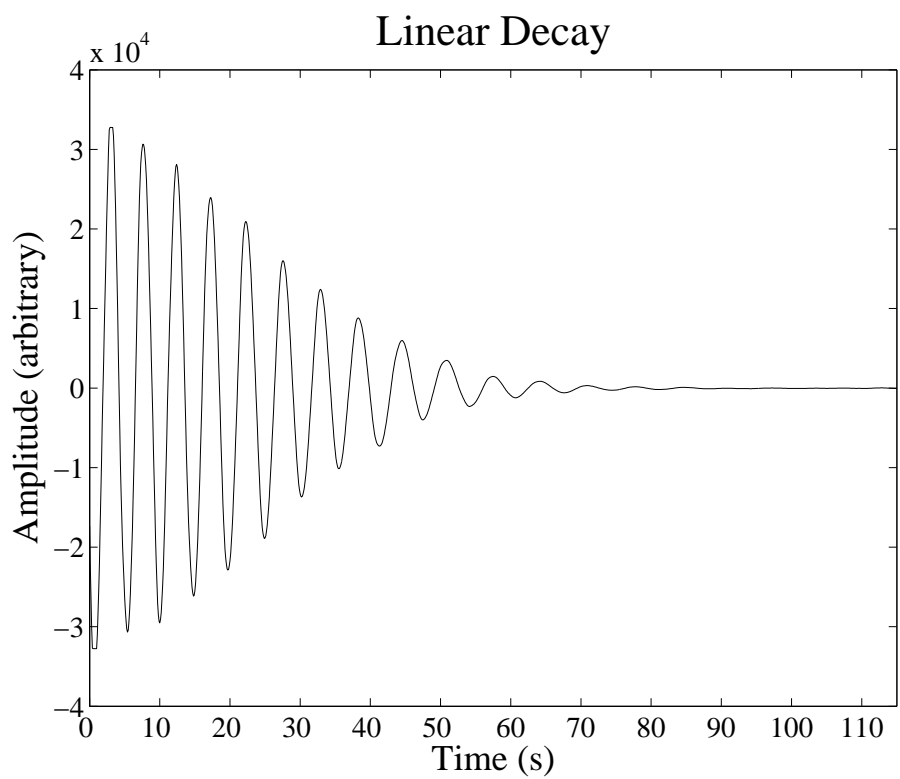}\vspace{2ex}
\caption{A typical modal ringdown when the sample is limited by 
the charge-correlated loss mechanism.  The envelope is linear in 
shape.
}\label{fig:linringdown}
\end{figure}

Finally, we gave the sample a uniform charge and varied the distance between the exciter
and sample.  No change in the ringdown was observed until the exciter came very close
to the sample, within about 300~$\mu$m.  Within this distance, two changes were observed;
the ringdowns went from an exponential shape to linear, and the characteristic time 
for a ringdown dropped precipitously.   A graph of a typical linear ringdown from when
the exciter was very close to the sample is shown in 
Figure~\ref{fig:linringdown}.
The change in ringdown versus distance between the sample and the exciter inside this
close region was investigated.  
 
\subsection{Results}

In order to investigate eddy current damping, polarization loss, and coupling to a lossy 
mechanical system, all discussed in Section~\ref{section:theory}, we measured Q 
versus charge density.  The results, shown in Table~\ref{table:QvsCharge}, indicate that the 
mechanical loss in silica is unaffected directly by charge on the surface in our experimental
setup.

To test the hypothesis that the charged sample was coupled to the lossy exciter structure, we 
tried varying the DC voltage on the exciter during the ringdown of a charged sample. The 
results shown in Table~\ref{table:QvsStatic} indicate 
that this mechanism did not contribute to the loss in our setup.

The only trial that gave results different from the uncharged sample was with the 
exciter extremely close to the charged sample.
Based on the linear shape of these ringdowns, Coulomb damping from dust
rubbing against the sample, discussed in Section~\ref{rubbing}, is the 
best fit of all the models discussed in Section~\ref{section:theory}.  
None of the other models predict this
linear decay shape.  All of these other models are consistent with no observable effect for some 
reasonable collection of parameters. 

\begin{figure}
\includegraphics[width=12cm]{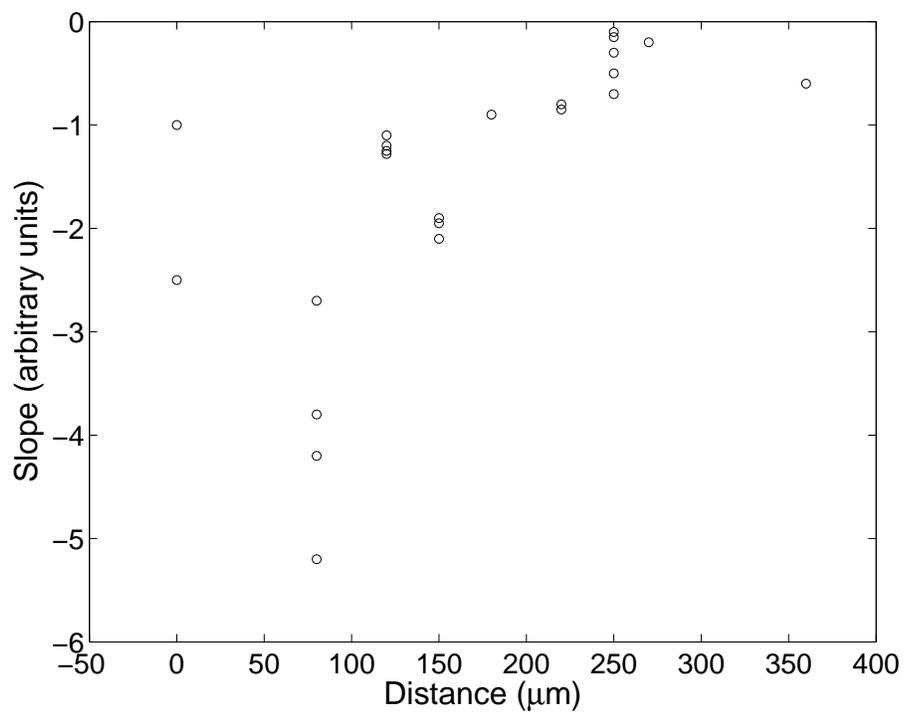}\vspace{2ex}
\caption{The slope of the linear envelope of the decay versus distance 
from the sample.  Zero distance is arbitrarily set as the closest point at 
which data was taken. There is a correlation between slopes closer to 0 and
greater distance between the sample and the exciter.
}\label{fig:slopevsd}
\end{figure}

The Coulomb damping model predicts that the slope of the linear decay envelope should
follow Eq.~(\ref{eqn:linear}).  This equation says that the slope gets steeper
as the normal force between the sample and the exciter increases.  The normal force between
the dust and the sample is caused by either the spring constant of the dust itself, or
of the pendulum suspension of the sample.  As the exciter is moved closer to the sample,
the normal force will increase in either case.  The slope of the linear ringdowns versus distance 
between the exciter and the sample is plotted in Figure~\ref{fig:slopevsd}.  This figure
suggests a correlation between low slope values and distance, in agreement with Eq.~(\ref{eqn:linear}).

\section{Implications}

The silica test masses currently installed in the LIGO vacuum 
system are known to become charged.  Measurement, control, and mechanical structures lie
within serveral millimeters of the surface of the suspended optics.  This combination of a 
charged sample and relatively close proximity to
a separate body could allow for Coulomb damping from dust spanning the gap.
In our laboratory experiments we never observed any Coulomb damping when the distance between the
exciter and the sample was this large, however.  The LIGO
vacuum chambers are always surrounded by class 100 portable clean rooms 
whenever the chambers are opened.  It is much less likely that dust could
contaminate the LIGO optics than the sample we measured in an open laboratory.
It is unlikely that conditions similar to what we experienced in the laboratory would allow 
for dust to cause Coulomb damping on the LIGO test masses.  

The vacuum chamber walls and the metal support structures around installed LIGO optics could act
as ground planes for charged optics.  None of these conductors are close enough to any optic for the
polarization losses discussed in Section~\ref{section:polarization} to be important.  The chamber 
walls are meters away while the support structure is tens of centimeters.

We did not investigate, either theoretically or experimentally, any noise sources
beyond excess thermal noise that could be caused by charged optics. It is possible that patchy 
charge densities could have thermally driven fluctuations.  Interactions with nearby ground planes
or background electric fields could then cause noise in the interferometer.  This possibility may 
require further study if charging of optics continues to be a problem.

\section*{Acknowledgments}
We would like to thank Mike Zucker and Myron MacInnis for help with our 
vacuum system, Doug Cook at the LIGO Hanford Observatory for loaning us 
the electrometer, and Rai Weiss, Peter Fritschel, and Phil Willems for 
useful comments and discussion. The LIGO Observatories were constructed by
the California Institute of Technology and Massachusetts Institute of 
Technology with funding from the National Science Foundation under  
cooperative agreement PHY-0107417.  This paper has been assigned LIGO
Document Number LIGO-P020014-00-D.
 
 \pagebreak

\pagebreak

\begin{table}
\begin{tabular}{cc}
{~Charge Density~($\mu$C/m$^2$)} & {~Measured Q~}\\
\hline
30  & $12.9 \times 10^6$ \\
90  & $12.9 \times 10^6$ \\
200  & $13.1 \times 10^6$ \\
\end{tabular}
\caption{Measured $Q$'s for varying surface charge densities on the silica disk.
The charge density was calculated from a measured electric field.}
\label{table:QvsCharge}
\end{table}

\begin{table}
\begin{tabular}{cc}
{~Exciter DC Potential (V)~} & {~Measured Q~}\\
\hline
250  & $14.3 \times 10^6$ \\
500  & $14.9 \times 10^6$ \\
1000  & $14.5 \times 10^6$ \\
\end{tabular}
\caption{Measured $Q$'s for varying DC Voltages on the exciter while the silica
disk was charged to 30 $\mu$C/m$^2$.}
\label{table:QvsStatic}
\end{table}

\end{document}